\documentclass[aps,pra,preprint,groupedaddress]{revtex4}

\usepackage{graphicx}
 
\newcommand{\be}{\begin{equation}}
\newcommand{\ee}{\end{equation}}

\begin{document}

\title{Wetting transition of water on graphite and other surfaces }

\author{ S.M. Gatica$^1$ J. K. Johnson$^{2,3}$, X.C. Zhao$^2$, and  M.W. Cole$^1$}

\affiliation{
$^1$  Department  of
Physics, Pennsylvania State
University, University Park, PA 16802.\\
$^2$ Department of Chemical and Petroleum Engineering, University of Pittsburgh, Pittsburgh, PA 15261.\\
$^3$ National Energy Technology Laboratory, US Department of Energy, P O Box 10940, Pittsburgh, PA 15236.}

\date{\today}

\begin{abstract}

A wetting transition occurs when the contact angle of a liquid drop on a surface changes from a nonzero value to zero. Such a transition has never been observed for water on any solid surface. This paper discusses the value of the temperature $T_w$ at which the transition should occur for water on graphite. A simple model, previously used for nonpolar fluids, predicts the value 
of $T_w$ as a function of the well-depth $D$ of the adsorption potential. While $D$ is not well known for the case of water/graphite, the model implies that $T_w$ is likely to fall in the range 350 to 500 K. Experimental search for this transition is warranted. Water wetting transition temperatures on other surfaces are also discussed. 

\end{abstract}

\maketitle

\section{Introduction}

Fundamental discoveries related to the thermodynamic laws 
describing adsorbed films on planar surfaces are still being made, 
despite the fact that these laws have been known for more than a century. 
An important example involves the contact angle $\theta$ 
of a liquid drop on a surface and its 
relevance to wetting transitions \cite{RowlinsonW89,Davis96,Dietrich88,RossB01}. Young's equation,
\be
\sigma_{gs} = \sigma_{ls}   + \sigma_{lg} \cos(\theta),\nonumber
\label{eq:YoungsEq}
\ee	
implies that $\theta$ vanishes when a particular equality is satisfied:	 
\be
\sigma_{gs} = \sigma_{ls}   + \sigma_{lg}.
\label{eq:surface-tension1}
\ee		
Here, $\sigma_{gs} $, $\sigma_{ls}$ and  $\sigma_{lg}$ are the gas-solid, liquid-solid, and liquid-gas interfacial tensions, respectively. This description in terms of macroscopic surface tensions applies only at saturated vapor pressure (svp). A wetting transition occurs when $\theta$ changes from a nonzero value to zero \cite{footnote1,RagilMBIB96}. Analyses describing this transition were presented some 25 years ago in complementary studies of Cahn and Ebner and Saam \cite{Cahn77,EbnerS77}. Their research found that if a fluid does not wet a particular surface at low temperature, then a wetting transition is expected to occur at some higher temperature $T_w$ that is less than the bulk critical temperature $T_c$;  $T_w$ thus separates nonwetting from wetting regimes. A drying transition ($\theta=\pi$), hypothesized in the early work, is believed now to be ruled out in the prevalent case of long range van der Waals forces \cite{EbnerS87,AncilottoCTC01}. 
 
Such wetting transitions were first observed some ten years ago in 
experiments involving simple gases (He, Ne and H$_2$) on alkali metal 
surfaces \cite{RossTR95,RossPRT97,Bojanetal99,MisturaLC94,RossTR98,Hallock95,HessSC97,DemolderBND95,WyattKS95}; 
soon afterwards, transitions were found with Hg on the surfaces of Ta and sapphire \cite{YaoH96,HenselY98,OhmasaKY98,OhmasaKY01,Kozhevnikovetal97,KozhevnikovANF98}. The common origin of these phenomena is that the gas-surface interaction $V(r)$ is only weakly attractive (implying nonwetting behavior at temperature $T$ close to the triple temperature $T_3$) \cite{Bojanetal99,FinnM89,ChizmeshyaCZ98,AncilottoFT00}. The dividing line between wetting and nonwetting behaviors at $T_3$ corresponds approximately to the situation when the adsorption well-depth $D= 3.5\epsilon$ \cite{CurtaroloSBCS00}, where $\epsilon$ is the well-depth of the adsorbate's intermolecular interaction; for a monatomic, classical fluid, this criterion is equivalent to the condition $D/(k_B T) \sim 6$  at the triple point \cite{footnote2}. The physics of this phenomenon is qualitatively simple: at pressures close to svp the gas is on the verge of condensing to form a bulk liquid phase. If the surface provides a very strongly attractive interaction, it nucleates the film, which grows continuously on the surface as the pressure $P$ approaches its value $P_0$ at svp. If, instead, the interaction is only weakly attractive, no such nucleation occurs below svp. In that case, a fluid droplet on the surface at svp will bead up instead of spreading across the surface. The difference between the two kinds of behavior is thus associated with the relative strength of cohesive and adhesive interactions. 
 
Interestingly, no wetting transition involving water has ever been observed, to the best of our knowledge \cite{footnote3,Wongetal97}. This situation is somewhat surprising in view of the fact that water is a much-studied fluid that does not wet many surfaces; transitions are expected to occur for each of these cases. This paper is concerned primarily with the wetting behavior of water on graphite. There exist several reasons for this choice of substrate. One is that flat surfaces of graphite, with few impurities, are readily available in a number of forms. A second is that a large database exists for other films on this surface \cite{BruchCZ97,ShrimptonCSC92,Zeppenfeld01}. In addition, the problem of water adsorption on graphite is related to that of water on and inside carbon nanotubes, a subject of considerable current interest \cite{GordilloM03,MaibaumC03,RiveraMC02,Werderetal01}. Finally, and most importantly, water is known to not wet the surface of graphite at room temperature; different values of the contact angle have been reported, presumably due to surface imperfections \cite{LunaCB91,MiuraM91,Scharder80,Morcos72,FowkesH40}. Thus, the predicted value of $T_w$ is an open question.  
 
The outline of this paper is the following: The next section describes a 
simple model of the wetting transition, used previously with much success 
in predicting transition behavior of both classical and quantum fluids. 
Section~\ref{sec:Potential} discusses calculations of the potential and the resulting 
values of $T_w$ for water/graphite.
Analogous values of $T_w$ for water on other surfaces are also discussed.
Section~\ref{sec:Summary} summarizes our results and 
conclusions. 
 
\section{Simple Model of the Transition }

In a 1991 paper, Cheng et al.\ \cite{ChengCST91} presented a so-called ``simple model" that derives predictions of $T_w$ from the adsorption potential and the $T$-dependent values of the adsorbate's density  $\rho$ and $\sigma_{lg}$. It achieves this goal by making a drastic assumption about the difference between two of the tensions appearing in Young's equation: 
 
\be
\sigma_{ls}   - \sigma_{gs} = \sigma_{lg} + \rho \int dz V(z)		
\label{eq:surface-tension2}
\ee
The logic of this relation is that $ \sigma_{ls} $   includes two contributions: the free energy associated with terminating the liquid (roughly equal to $\sigma_{lg} $) and the liquid-surface interaction energy. In eq~\ref{eq:surface-tension2}, the domain of integration extends between the minimum in the adsorption potential (at $z=z_{min}$ ) and infinity. The qualitative justification for this relation includes some implicit assumptions: (a) that a very low density film is present on the surface just below the transition pressure, so that $\sigma_{gs}$ is essentially the surface tension of the bare substrate, (b) that when an infinitely thick liquid is in contact with the substrate its excess free energy differs from that of the bulk liquid-vapor interface by the integrated potential energy, (c) that the function $V(r)$ is assumed to depend on just the normal coordinate $z$ and (d) that the integrated potential energy is adequately described by a ``sharp kink" approximation, in which the fluid density rises from zero to its bulk value at position $z= z_{min}$. If eq~\ref{eq:surface-tension2} is valid, then the wetting transition relation, eq~\ref{eq:surface-tension1}, can be manipulated to yield the following equation: 
\be
\left(\frac{2 \,\sigma_{lg}}{\rho}\right)_T = - \int dz V(z)	= I.	
\label{eq:simple-theory}
\ee 
Equations~\ref{eq:surface-tension2} and \ref{eq:simple-theory} provide an implicit prediction for $T_w$, the temperature for which the left side equals the integral $I$, an explicit function of the adsorption system. The accuracy of eq~\ref{eq:simple-theory} can be tested by experiments, if the potential is accurately known, or by simulations, if it is not, by using a hypothetical potential. An extensive set of classical simulations were undertaken by Curtarolo et al., who tested the model by comparing its predictions for wetting at the triple point with simulation data for a wide variety of adsorption potentials \cite{CurtaroloSBCS00}. The results revealed that the model works well, overall, except for situations involving the very least attractive interactions, in which case the error was not negligible. More recently, path integral Monte Carlo simulations were undertaken by Shi et al.\ \cite{ShiJC03}. They found the simple model's predictions of $T_w$ to differ from the exact results by about 10\% (near 20 K) for the isotopes of hydrogen on several alkali metal surfaces. Comparison with experimental data indicates in most cases that eq~\ref{eq:simple-theory} predicts $T_w$ relatively well when the potential is accurately known\cite{RossTR95,RossPRT97,Bojanetal99,MisturaLC94,RossTR98,Hallock95,HessSC97,DemolderBND95,WyattKS95}. 

One model potential that has been used extensively to treat the wetting problem is: 
\be
V(z)= \frac{4 \,C_3^3}{27\, D^2\,   z^9}- \frac{C_3}{z^3}.	
\label{eq:3-9potential}
\ee
This 3-9 potential is analogous to the Lennard-Jones 6-12 interatomic potential and is usually adopted for similar reasons; it combines a rigorously correct form of the asymptotic attraction ($V\sim  C_3/z^3$) and a simple power law repulsion to yield a qualitatively plausible and mathematically convenient functional form. The asymptotic van der Waals coefficient $C_3$ is discussed in some detail in the next section. If one inserts eq~\ref{eq:3-9potential} into eq~\ref{eq:simple-theory}, with  
$z_{min}= \left[2 \,C_3/(3D)\right]^{1/3}$ , one obtains 
\be
I = \frac{11}{24} \left(\frac{3}{2}\right)^{2/3} \left(C_3 \,D^2  \right)^{1/3}.  
\label{eq:Integral}
\ee 
When combined with eq~\ref{eq:simple-theory}, this equation yields a simple relation that we employ to determine $T_w$: 
\be
\left(C_3 \,D^2  \right)^{1/3} = 3.33 \left(\frac{\sigma_{lg}}{\rho}\right)_T. 	
\label{eq:C3Tw}
\ee 
We emphasize that this relation is derived by combining the simple model of the wetting transition with the simple 3-9 model potential. In spite of these approximations, the result appears to describe most of the wetting studies carried out so far, at least semiquantitatively. In the next section, we evaluate the predictions of the simple model for both the 3-9 potential and other water/graphite potentials. 
 
\section{The Adsorption Potential }
\label{sec:Potential} 

\subsection{Dispersion interaction }
 
The first aspect of the adsorption potential considered here is the van der Waals dispersion coefficient $C_3$. While values of $C_3$ exist for many gas/surface combinations \cite{VidaliIKC91}, no values have been calculated previously for the particular case of water/graphite. In the case of a molecule, there exist two independent contributions to this coefficient: 
\be 
C_3 = C_{\mathrm{perm}} + C_{\mathrm{vdw}}. 		
\label{eq:C3potential}
\ee
The first term arises from the interaction between the molecule's dipole moment ${\bf p}$ and its image in the dielectric substrate, while the second term is due to the coupled dipolar charge fluctuations of the electrons on the molecule and those of the substrate \cite{BruchCZ97}. In computing both of these terms, we assume a Drude model form for the surface dielectric response function: 
\be
\frac{\varepsilon(i E) -1}{ \varepsilon(i E) +1}=\frac{g}{1+(E/E_s)^2}.	
\label{eq:Drude}
\ee 
Here, $\varepsilon(i E)$ is the dielectric function at imaginary energy $iE$ and $E_s$ is a characteristic electronic energy of the solid. This form is exact, with $g=1$ and $E_s$ equal to the surface plasmon energy, for a perfect conductor. It is also exact for a low density solid comprised of Drude oscillators, representing the individual molecules. For many other materials, this parameterization has been shown to work well in determining the various dispersion coefficients \cite{RauberKCB82}.  
 
In evaluating $C_{\mathrm{perm}}$, we adapt the derivation of London \cite{London30a,London30b} to the case of a molecule having a permanent dipole moment $\bf p$ oriented at an angle $\phi$ relative to the surface normal, so that $p_z=p \,\cos\phi$, taking into account the dielectric screening at low energy (relevant to the slow rotation of $\bf p$): 
\be 
C_{\mathrm{perm}}(\phi)= \frac{g \,p^2(1+\cos^2\phi)}{16}.		
\label{eq:Cperm1}
 \ee
In the case of a randomly oriented, or freely rotating molecule, one obtains 
\be 
C_{\mathrm{perm}} =\langle C_{\mathrm{perm}}(\phi)\rangle= \frac{g \,p^2}{12},
\label{eq:Cperm2}
\ee 
since the average of $\cos^2\phi$ for a randomly oriented dipole is 1/3. 
In the case of water/graphite, $g=0.619$ \cite{BruchCZ97} and $p=1.85$ Debye \cite{CloughBKR73}, so that (in this random case)  
$C_{\mathrm{perm}}$ =104 meV-\AA$^3$	

As to the van der Waals part of the interaction, an analogous modeling of the dynamical polarizability of water (with a characteristic energy $E_{\mathrm{water}}$) yields an expression 
\be 
C_{\mathrm{vdw}}= \frac{g\, \alpha\, E_s}{8\, (1+\frac{E_s}{E_{\mathrm{water}}})},		
\label{eq:Cvdw}
\ee
Using data from Appendix E of Bruch et al.\ \cite{BruchCZ97}, with $E_{\mathrm{water}}$ =18.1 eV and a rotationally averaged polarizability $\alpha$= 1.4 \AA$^3$, we obtain a result $C_{\mathrm{vdw}}$ = 971 meV-\AA$^3$ . From eq~\ref{eq:C3potential}, the total $C_3$ =  1075 meV-\AA$^3$. Note that the permanent moment's contribution to this coefficient is about 10\%. 
 
\begin{figure}
\centerline{\includegraphics[width=0.7\textwidth]{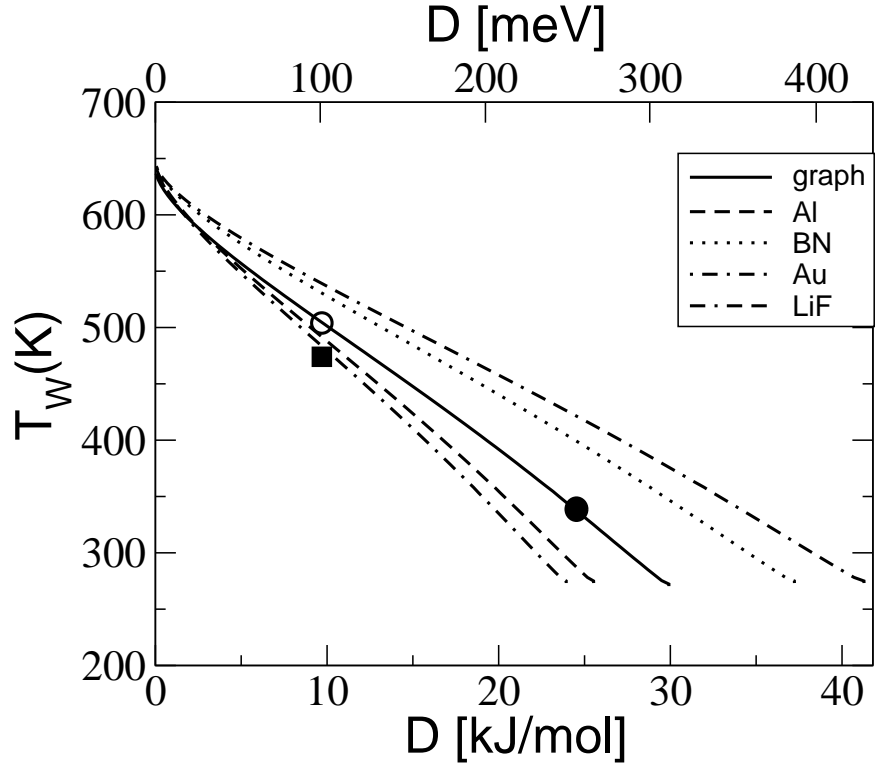}}
\caption{Wetting temperature of water as a function of  the well-depth $D$ of the adsorption potential on graphite (full curve), using eq~\ref{eq:C3Tw} and $C_3$ =  1075 meV-\AA$^3$. The  square is obtained from integration of the ZJ 
potential, eq~\ref{eq:new-pot}, as described in the text. The open circle is based on the well-depth used by Gordillo and Mart\' {\i} \cite{GordilloM03}, while the filled circle uses the calculated value of Feller and Jordan \cite{FellerJ00}. Other curves are dependences computed for the indicated surfaces. 
}
\label{fig:TwD}
\end{figure}

Using this value of $C_3$ in eq~\ref{eq:C3Tw}, we 
can predict $T_w$ as a function of the well depth, $D$. 
In Figure~\ref{fig:TwD} we plot $T_w$ as a function of $D$. 
The surface tension and density values for water used to 
construct Figure~\ref{fig:TwD} 
were taken from Ref. \cite{NISTWebBook}. 

\subsection{Well-depth calculations }
 
There have been numerous theoretical and semiempirical studies of the 
water-graphite interaction, as have recently been discussed by Werder et al.\ 
and Pertsin and Grunze \cite{WerderWJHK03,PertsinG03}. The values emerging from these 
studies vary between 7.1 and 24.3 kJ/mole. This range reflects the wide 
variety of methods and assumptions employed in these calculations. 
Figure~\ref{fig:TwD} indicates the corresponding values of the wetting 
temperature derived from some of these well-depths. 
 
\subsection{ Results for a new potential }
 
Each carbon atom in graphite has a local quadrupole 
moment due to the symmetry of the crystal \cite{NicholsonCP90,Cracknell90,VernovS92,WhitehouseB93,HansenBR92}.
Vernov and Steele developed an angle-explicit, position-dependent potential 
that accounts for the interactions of a polar molecule with the quadrupoles 
on the carbon atoms of graphite \cite{VernovS92}. They found that the 
corrugation of the water-graphite potential energy surface in the $x,y$ plane 
was dramatically enhanced by the inclusion of the dipole-quadrupole term. 
Zhao and Johnson have recently developed approximate 
expressions for the dipole-induced dipole, dipole-quadrupole, and 
quadrupole-quadrupole potential terms for polar fluid molecules 
interacting with graphite \cite{ZhaoJ04}.
The expressions are angle averaged and integrated, so that 
they depend only on the normal coordinate, $z$. 
For water on graphite only the 
dipole-induced dipole and dipole-quadrupole terms are important. These 
are given by
\begin{equation}
V_{\mathrm{polar}} = 
\frac{\Delta\rho_{\mathrm{s}}\pi p^{2}}
{(4\pi\varepsilon_{0})^{2}}
\left\{
\frac{\alpha_{\mathrm{c}}}
{2}
\left[\frac{1}{z^4}+\frac{1}{3\Delta(z+\Delta)^{3}}\right] 
+
\frac{\Theta_{\mathrm{c}}^2}
{3k_BT}
\left[\frac{1}{z^6}+\frac{1}{5\Delta(z+\Delta)^{5}}\right]
\right\},
\label{eq:new-pot}
\end{equation}
where $\Delta$ is the distance between the graphene layers, 
$\rho_{\mathrm{s}}$ is the density of atoms in graphite, 
$\varepsilon_0$ is the vacuum permittivity, $p$ is 
the magnitude of the dipole moment of water, $\alpha_{\mathrm{c}}$ is the 
angle averaged polarizability of a carbon
atom in graphite, and  $\Theta_{\mathrm{c}}$ is the quadrupole moment on each
carbon atom in graphite. The polar terms in eq~\ref{eq:new-pot} are 
added to the ``Steele model" 10-4-3 potential \cite{Steele73,Steele93} 
to obtain the full potential. 

\begin{figure}
\centerline{\includegraphics[width=0.7\textwidth]{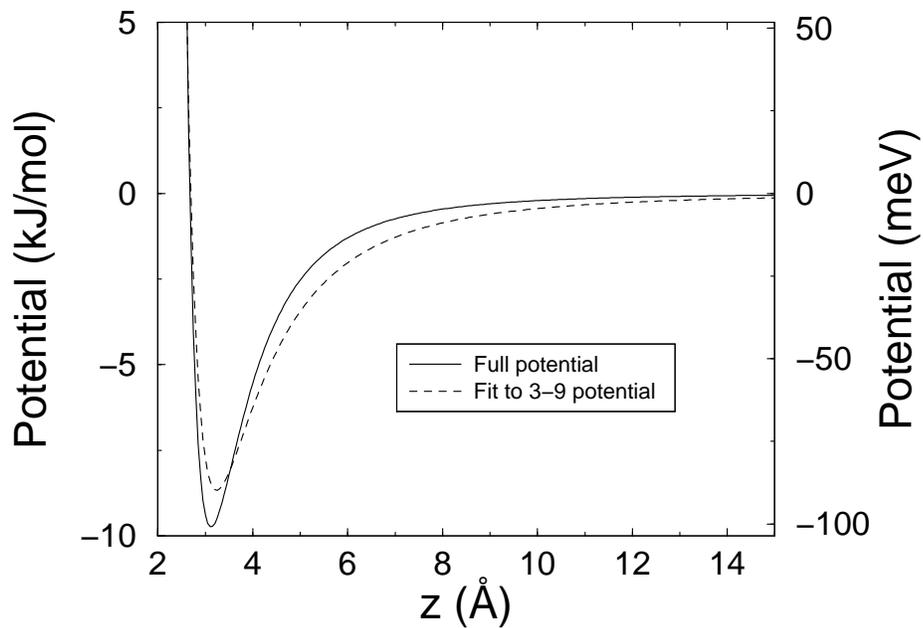}}
\caption{Potential energy function of Zhao and Johnson \cite{ZhaoJ04}, 
discussed in the text, evaluated at $T=474$ K (full curve). The dashed curve is a 3-9 potential fit to this potential.  
}
\label{fig:potentials}
\end{figure}

The numerical value of the potential depends on the specific fluid 
potential used for water. In this work we employ the Lennard-Jones 
parameters from the 
TIP4P potential \cite{Jorgensenetal83} for calculating the 10-4-3 van der Waals solid-fluid 
contribution. For the polar terms, we use a point dipole having the 
experimental gas phase value of 1.85 Debye. The potential depends on 
temperature, due to Boltzmann angle averaging of the dipole-quadrupole term. 
However, the temperature dependence is relatively weak. The water-graphite 
potential at 474 K and a fit of this potential to the 3-9 functional form, 
eq~\ref{eq:3-9potential}, are plotted in Figure~\ref{fig:potentials}. 
The the main difference between the 
Zhao-Johnson (ZJ) potential and the 3-9 potential 
developed above is that the ZJ potential accounts for 
dipole-quadrupole interactions but calculation of the 
$C_3$ term does not. The dipole-quadrupole term accounts 
for 12\% of the energy at the potential minimum at 474 K, whereas 
the induction term is 22\%. 
Note that the 3-9 functional form does not accurately describe the 
ZJ potential. The 3-9 potential is not as deep as the full 
potential, but it has a longer range (decays more slowly). 
 
With this new potential, one can evaluate the simple model to predict $T_w$ 
in three ways. One is by directly integrating $V(z)$, including its weak 
$T$-dependence, and using eq~\ref{eq:simple-theory}. A second is to employ the simplified 
expression in eq~\ref{eq:C3Tw}, using a 3-9 potential fit to the theoretical potential; 
the fit values are $D=89.8$ meV and $C_3 = 4751$ meV-\AA$^3$. 
 A third way to predict T$_w$ is to use the 
theoretical well-depth (9.74 kJ/mol or 101 meV) and the theoretical 
value of $C_3$ , i.e., use the curve in Figure~\ref{fig:TwD}. 
The results in these three cases are 
$T_w=474$, 416, and 504 K, respectively. Of these three, the value 
$T_w =474$ K, obtained from the direct integration, is the recommended 
choice since it is based on the theoretical potential, without recourse 
to a preconceived model potential. The 3-9 fit and resulting low value of 
$T_w$ are particularly suspect because the fitted value of $C_3$ is a 
factor of four greater than the value ($C_3=1075$ meV-\AA$^3$ ) derived 
in the previous section, which has an estimated uncertainty of 20 \% 
(based on previous studies of $C_3$ values \cite{VidaliIKC91}). 
The third value coincides with the one predicted from  the well depth calculated by Gordillo et al. \cite{GordilloM03}

\subsection{ Wetting on other surfaces }
 
Similar calculations of $C_3$ have been carried out for water on other surfaces, with results presented in Table 1. For any one of these surfaces, say $x$, one can compute the dependence $D_x(T_w)$ of the well depth on $T_w$ from eq~\ref{eq:C3Tw}, by scaling from the functional dependence $D_g(T_w)$ for graphite, shown in Figure~\ref{fig:TwD}: 

\be 
D_x(T_w)= \left(\frac{C_g}{C_x}\right)^{1/2}\, D_g(T_w). 
\label{eq:Dx}
\ee 
Here, $C_x/C_g$ is the ratio of van der Waals coefficients for the surface of interest to that on graphite. 
Figure~\ref{fig:TwD} shows this behavior for these various surfaces.

\section{Summary and Conclusions}
\label{sec:Summary}

Although water does not wet many surfaces, no wetting transition for 
water has been observed. Quite general arguments predict such a 
transition, unless the adsorption potential is extremely weak 
(as occurs for Ne/Cs \cite{AncilottoCTC01}). In this paper we have predicted 
the wetting temperature as a function of $D$, using a simple model 
that has been found reliable for spherical adsorbates \cite{CurtaroloSBCS00}. 
One reason for concern about the reliability of the prediction is that the method has not yet been tested by comparing with results for water {\em per se}. Water is greatly affected by electrostatic forces absent from the nonpolar systems. However, the simple physical picture underlying the model and its previous success both suggest that it should be quite useful in motivating and guiding experimental searches for such a transition. 
 
Evidently, this subject can benefit from further research in three related directions. One is to carry out simulations of water/graphite based on reliable intermolecular interactions in order to determine the accuracy of the simple model employed here. A second is to further refine the water/graphite interaction potential, which is very uncertain. A third is an experimental search for the predicted wetting transition. 
 
We acknowledge very helpful discussions with Ken Jordan and Erwin Vogler. We are grateful for support from NSF (grants 02-08520 and 03-03916 and EEC 0085480 (JKJ)).

\newpage

\begin{table}[tbh]
\begin{tabular}{|c|c|c|c|c|c|} \hline
 
           &		graph	&BN	&Al	&Au	&LiF \\ \hline

 $g$ 	&		0.619 &	0.38 &	0.98 &	0.84 &	0.31 \\
 
$E_s$ (au) 	&	0.667 &	0.71 &	0.473 &	0.888 &	0.74 \\
 
$C_{\mathrm{perm}}$ (meV-\AA$^3$) & 	104 &	64 &	166 &	142 &	51 \\
 
$C_{\mathrm{vdw}}$ (meV-\AA$^3$)	&	971 &	614 &	1278 &	1502 &	511 \\
 
$C_{\mathrm{tot}}$(meV-\AA$^3$)	&	 1074 &	678 &	1444 &	1644 &	562 \\ \hline
 
\end{tabular}
\caption{van der Waals interaction coefficient and other parameters relevant to wetting of water on indicated surfaces. Data for $g$ and $E_s$ are taken from Bruch et al \cite{BruchCZ97}, while the other quantities are computed from equations in the text. }

\end{table}

\bibliographystyle{jpc}
\bibliography{waterbib}

\begin{thebibliography}{10}
\bibitem{footnote1}
Thus far, only first order transitions have been observed on solid surfaces,
  but higher order transitions have been observed on liquid surfaces; see ref.
  \protect\cite{RagilMBIB96}.
\bibitem{footnote2}
We note that the gases Ne, Ar, Kr and Xe have the ratio
  $\varepsilon/k_BT_3=1.7$ approximately where $\varepsilon$ values are well
  depths from experimental data. (See Table A2 in Bruch et
  al.\cite{BruchCZ97}).
\bibitem{footnote3}
A different kind of surface wetting transition has been attributed to the
  presence of water on the surface of caprolactam. See ref.~\cite{Wongetal97}.
\end{thebibliography}

\newpage

\end{document}